\documentclass[ reprint,superscriptaddress, amsmath,amssymb, aps,nolongbibliography]{revtex4-1}

\usepackage[colorlinks=true,linkcolor=blue,urlcolor=blue,citecolor=blue]{hyperref}
\usepackage{overpic}
\usepackage{upgreek}
\usepackage{physics}
\usepackage{graphicx}
\usepackage{dcolumn}
\usepackage{bm}
\usepackage{outlines}
\usepackage[dvipsnames]{xcolor}
\usepackage{gensymb}

\usepackage{pdfpages}
\makeatletter
\AtBeginDocument{\let\LS@rot\@undefined}
\makeatother

\newcommand{\mitll}{MIT Lincoln Laboratory, Lexington, MA 02421, USA}
\newcommand{\mitNSE}{Department of Nuclear Science and Engineering, Massachusetts Institute of Technology, Cambridge, MA 02139, USA}
\newcommand{\mitRLE}{Research Laboratory of Electronics, Massachusetts Institute of Technology, Cambridge, MA 02139, USA}
\newcommand{\mitPhy}{Department of Physics, Massachusetts Institute of Technology, Cambridge, MA 02139, USA}

\begin{document}

\title{Quantum Frequency Mixing using an N-$V$ Diamond Microscope}

\author{Samuel J. Karlson}
\affiliation{\mitll}
\affiliation{\mitNSE}
\affiliation{\mitRLE}
\author{Pauli Kehayias}
\affiliation{\mitll}
\author{Jennifer M. Schloss}
\affiliation{\mitll}
\author{Guoqing Wang}
\affiliation{\mitNSE}
\affiliation{\mitRLE}
\affiliation{\mitPhy}
\author{Andrew C. Maccabe}
\altaffiliation{Current Affiliation: Quantum Science and Engineering Program, Harvard University, Cambridge, MA 02138, USA}
\affiliation{\mitll}
\author{Adam Libson}
\affiliation{\mitll}
\author{David F. Phillips}
\affiliation{\mitll}
\author{Paola Cappellaro}
\affiliation{\mitNSE}
\affiliation{\mitRLE}
\affiliation{\mitPhy}
\author{Danielle A. Braje}
\email{braje@ll.mit.edu}
\affiliation{\mitll}

\date{\today}

\begin{abstract}
Wide-field magnetic microscopy using nitrogen-vacancy (NV) centers in diamond can yield high-quality magnetic images of DC and AC magnetic fields. The unique combination of micron-scale spatial resolution of scalar or vector fields at room temperature and parallel camera readout make this an appealing technique for applications in biology, geology, condensed-matter physics, and electronics. However, while NV magnetic microscopy has achieved great success in these areas, historically the accessible frequency range has been limited. In this paper, we overcome this limitation by implementing the recently developed technique of quantum frequency mixing. With this approach,  we generate wide-field magnetic images of test structures driven by alternating currents up to 70 MHz, well outside the reach of DC and Rabi magnetometry methods. With further improvements, this approach could find utility in hyperspectral imaging for electronics power spectrum analysis, electronics diagnostics and troubleshooting, and quantum computing hardware validation.
\end{abstract}

\maketitle

\section{Introduction}
Wide-field magnetic microscopy using nitrogen-vacancy (NV) centers in diamond has gained attention for its value in diverse applications including geology, biology, condensed-matter physics, and electronics \cite{edlynQDMreview, tetienneQDMreview}. Unlike single-pixel scanning magnetic microscopes, NV-diamond-based microscopes can simultaneously image all pixels in a wide (mm-scale) field of view (FOV) without moving parts, and can reveal spatiotemporal dynamics of magnetic fields in samples under study. The technique offers micron-scale spatial resolution and the potential for operation over a broad frequency range, and is being applied to a range of interdisciplinary applications.

Despite the ongoing accomplishments and progress, most wide-field NV magnetic microscopes have been limited to measuring static or low-frequency ($\lesssim$ 1 kHz) magnetic fields, with few exceptions \cite{linh_widefield, devience_nuclSens, maletinskyMWimger, wrachtrupCaF2, dominikMRI}. This is due to the added complexity required for measuring MHz-frequency and GHz-frequency AC magnetic fields (including pulsed NV interrogation and gated optical readout) and the greater need for homogeneous laser, microwave (MW), bias magnetic field, and NV strain profiles across the FOV.   Moreover, these measurements are restricted to specific and often narrow frequency ranges. For example, AC magnetometry using pulsed dynamical decoupling is limited by MW $\pi$-pulse durations to typically $<$ 10 MHz \cite{rmpQuantumSensingReview}, while Rabi-oscillation  AC magnetometry senses in a narrow frequency range ($\sim$1 MHz) around the NV spin resonance frequency (usually near $\sim$3~GHz) \cite{scott_pT_RabiMag}. Though varying a strong bias magnetic field can tune (and extend) the Rabi-oscillation AC magnetometry frequency range, field uniformity requirements render this method  impractical for macroscopic NV ensembles. As a result, NV AC magnetic imaging demonstrations often sacrifice frequency range, FOV size, or both.

Here we demonstrate a pulsed diamond magnetic microscope with a $1.5 \times 1.5$ mm$^2$ FOV size, and image DC and AC magnetic fields up to 70 MHz. To our knowledge, this is the largest-area pulsed NV magnetic imaging demonstration, as previous experiments have shown $\sim$0.5 mm$^2$ at best \cite{maletinskyMWimger}. To enable wide-frequency sensitivity, we apply the quantum frequency mixing (QFM) approach recently demonstrated with small NV ensembles \cite{guoqingQFM} to our wide-field imager.  We validate the technique by imaging AC magnetic fields from currents in a straight-wire test structure patterned on a glass slide, where we observe the expected frequency dependence of detected signals. We then image magnetic fields from a second test structure (an Archimedean spiral), demonstrating agreement between DC and AC images. Finally, we explore the QFM magnetometry dynamic range under different signal frequencies and amplitudes, leading to additional analytical expressions of the expected behavior.   Combining high-performance wide-field NV-diamond microscopy with the QFM AC magnetometry technique opens the door to new magnetic microscopy applications, including high-resolution power spectrum analysis (PSA) imaging of integrated circuits, electronics failure analysis, validation of $\sim$1-100 MHz electrical components, and even diagnostics of quantum computing hardware (e.g.~ion trap chips) \cite{aprelScanMS, neoceraSpaceDomainRefl, NVionTrapFA}.

\begin{figure*}[hbt]
\begin{overpic}[width=1\textwidth]{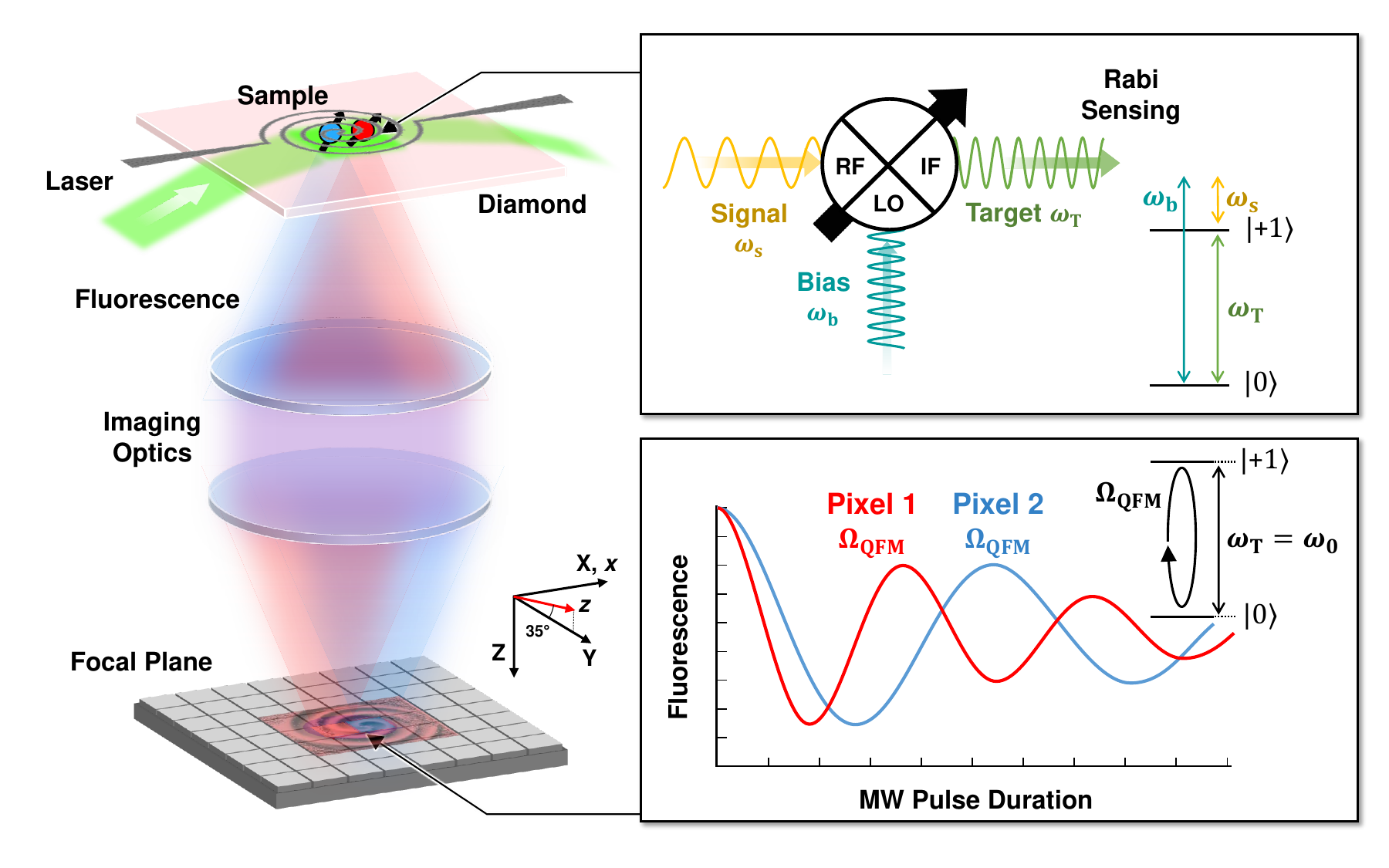}
\centering
\put(5,57){\textsf{\Large a)}}
\put(47,57){\textsf{\Large b)}}
\put(47,27){\textsf{\Large c)}}
\end{overpic}
\caption{\label{fig1drawing} 
(a) Schematic diagram of a QFM magnetic microscope. A green laser beam illuminates the NV layer. Different regions of the NV layer (highlighted by blue and red spins) respond to spatially varying magnetic fields and produce magnetic-field-dependent fluorescence, which is imaged onto a camera sensor using imaging optics. An example test structure (an Archimedean spiral), generating a spatially varying magnetic field, is shown at the sensor plane, and a magnetic map of its current is imaged at the focal plane. 
(b) The NV spins act as both quantum mixers and quantum sensors. In the presence of a signal field (yellow) at $\omega_{\mathrm{s}}$ and a bias field (teal) at $\omega_{\mathrm{b}}$, a target field (green) at the sum or difference frequency $\omega_{\mathrm{T}}$ is generated.  When $\omega_{\mathrm{b}}$ is chosen so that the sum or difference frequency is resonant with the NV resonance at $\omega_0$, population oscillations occur between the two spin states. 
(c) The oscillation frequency, called the target Rabi frequency $\Omega_{\mathrm{QFM}}$, provides a measure of the signal field amplitude $\Omega_{\mathrm{sz}}$, given by Eq.~\ref{eq:qfmConversion}. Spatial variations in $\Omega_{\mathrm{sz}}$ result in variations in $\Omega_{\mathrm{QFM}}$ from pixel to pixel, allowing wide-field imaging of AC signal amplitudes at arbitrary $\omega_{\mathrm{s}}$.
}
\end{figure*}

\section{Quantum Frequency Mixing}
Frequency mixing is a process that uses a nonlinear response to generate new frequencies from two periodic input signals. One example classical mixer, the unbalanced single-diode mixer, uses the nonlinear relationship between input voltage and response current to multiply two input signals, generating new frequencies at the sum and difference of the input frequencies. In our work using NV centers in diamond as a quantum frequency mixer, the NV quantum system is nonlinear, and we exploit this to multiply two AC magnetic signals within the quantum system itself \cite{guoqingQFM}. The type of mixing employed here is heterodyne, and the net effect of our experiment is implementation of a heterodyne receiver in a quantum system.

When a quantum system is driven with two fields at different frequencies, its dynamics can be described as a quantum frequency mixer, where the effective Hamiltonian derivation in the multi-mode Floquet picture allows us to identify the dynamics to describe a frequency-mixer-based quantum sensing protocol \cite{guoqingQFM}. Here, we summarize the principle of quantum frequency mixing and its application to this work.

To bring a signal of frequency $\omega_{\mathrm{s}}$ into resonance with an NV transition frequency $\omega_0$, we apply a bias oscillating field with a frequency $\omega_{\mathrm{b}}$. Here $\omega_{\mathrm{s}}$ and $\omega_{\mathrm{b}}$ are analogous to the input radio-frequency (RF) and local oscillator (LO) of a classical mixer. The effective signal generated by the nonlinear process of the quantum system is at frequency $\omega_{\mathrm{T}}$, which is analogous to the intermediate frequency (IF) of a classical mixer.

This process can be described by Floquet theory \cite{LESKES2010345,10.1063/1.2800319}. In Floquet analysis, the time-dependent Hamiltonian in the Hilbert space is converted to a time-independent Hamiltonian in Floquet space. The combined effect of two Fourier components $H_b e^{i\omega_{\mathrm{b}} t}$ and $H_s e^{i\omega_{\mathrm{s}} t}$ associated with the input signal $\omega_{\mathrm{s}}$ and the bias AC field $\omega_{\mathrm{b}}$, respectively, gives rise to a component at the target frequency $\omega_{\mathrm{T}}$. Upon block-diagonalization of the Floquet Hamiltonian, transforming back to the Hilbert space yields an effective Hamiltonian with a frequency $\omega_{\mathrm{b}}-\omega_{\mathrm{s}}$ \cite{guoqingQFM}. With an appropriate choice of $\omega_{\mathrm{b}}$, the effective target signal is now at the NV resonance frequency near 3 GHz, and existing quantum sensing protocols such as Rabi magnetometry can be used to probe the system.

The physics of a signal magnetic field projection along the NV axis, with frequency $\omega_{\mathrm{s}}$, detected using a Rabi oscillation measurement protocol (see Sec.~\ref{sec:ExperimentalProc} below) can be described with the Hamiltonian:
\begin{eqnarray}
H=\frac{\omega_0 }{2}\sigma_z +\Omega_{\mathrm{b}} \cos \left(\omega_{\mathrm{b}} t+\phi_\mathrm{b} \right)\sigma_x \nonumber \\+~\Omega_{\mathrm{sz}} \cos \left(\omega_{\mathrm{s}} t+\phi_\mathrm{s} \right)\sigma_z,
\label{eq:one}
\end{eqnarray}
where $\Omega_{\mathrm{b}}$ is the amplitude of the bias field and $\Omega_{\mathrm{sz}}$ is the amplitude of the signal frequency projected along the NV axis. In a rotating frame defined by $U=e^{-i\left(\omega_0 t/2\right)\sigma_z }$, the effective Hamiltonian contains two oscillating frequencies. When one is resonant with the NV transition, and the other is far-detuned and can be neglected in the spirit of the rotating-wave approximation.
We choose a bias frequency  $\omega_{\mathrm{b}}$ such that
\begin{eqnarray}
\omega_{\mathrm{b}}-\omega_0-\omega_{\mathrm{s}}= \delta_z ,
\end{eqnarray}
where $\omega_0$ is the NV resonance frequency and $\delta_z$ is a small detuning which we try to minimize in the experiment. Assuming $\Omega_{\mathrm{sz}} ,\Omega_{\mathrm{b}} ,\ll \omega_{\mathrm{s}}$, the amplitude of the effective target Rabi frequency is \cite{guoqingQFM}
\begin{eqnarray}
\Omega_{\mathrm{QFM}} =\frac{\Omega_{\mathrm{sz}} \Omega_{\mathrm{b}} }{\omega_{\mathrm{s}} }.
\label{eq:qfmConversion}
\end{eqnarray}
Table \ref{tbl:params} lists the relevant parameters for this experiment.

\begin{figure*}[hbt!]
\begin{overpic}[width=0.95\textwidth]{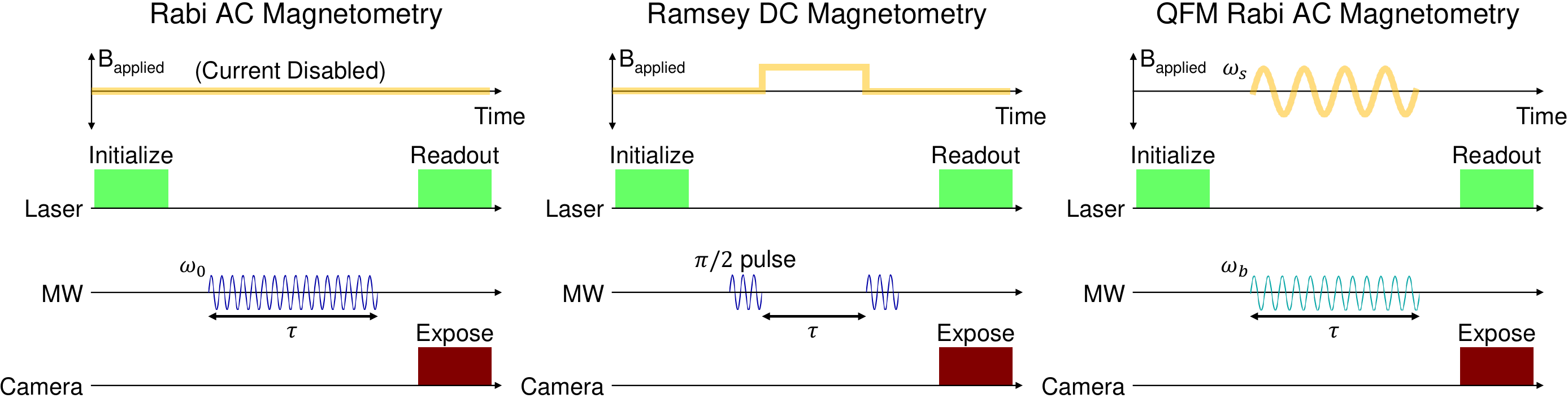}
\centering
\put(2,25){\textsf{\Large a)}}
\put(35,25){\textsf{\Large b)}}
\put(68,25){\textsf{\Large c)}}
\end{overpic}
\caption{
(a) Conventional Rabi pulse sequence used to determine the on-resonance NV Rabi frequencies, with no current applied to the test structure. We initialize the NVs to the $\ket{0}$ state with a laser pulse, apply an on-resonance MW field of frequency $\omega_0$  for a variable duration, and read out the NV final state with a second laser pulse. 
(b) Ramsey pulse sequence used for DC magnetic imaging. We apply a DC current to the test structure, and we apply a $\pi/2$ MW pulse to create a superposition of the  $\ket{0}$ and  $\ket{+1}$ states and allow the system to evolve for a fixed time $\tau$. We then apply a second $\pi/2$ pulse, followed by a laser readout pulse.
(c) Quantum frequency mixing (QFM) Rabi pulse sequence used to measure an AC magnetic field amplitude at frequency $\omega_{\mathrm{s}}$. This is similar to the conventional Rabi oscillation experiment, except with an applied bias frequency $\omega_{\mathrm{b}}$.}
\label{fig:PulseSequences}
\end{figure*}

\begin{table}[]
\begin{tabular}{c|c}
\textbf{Parameter} & \textbf{Description} \\ \hline
$\omega_0$         & NV resonance frequency    \\ 
$\Omega_0$         & On-resonance Rabi frequency    \\ \hline
$\omega_{\mathrm{b}}$         & Bias frequency       \\
$\Omega_{\mathrm{b}}$         & Bias amplitude       \\ \hline
$\omega_{\mathrm{s}}$         & Signal frequency          \\
$\Omega_{\mathrm{sz}}$         & Signal amplitude along $z$         \\ \hline
$\omega_{\mathrm{T}}$         & QFM target frequency            \\
$\Omega_{\mathrm{QFM}}$     & QFM Rabi frequency        \\
\end{tabular}
\caption{\label{tbl:params}
A list of relevant parameters describing the QFM AC magnetometry protocol.
}
\end{table}

\section{Experimental Methods}

\subsection{Microscope Apparatus}

Figure \ref{fig1drawing}a shows a schematic of the microscope apparatus. This drawing shows the relevant coordinate systems: $\{x,y,z\}$ is the NV coordinate system (with $z$ along the N-V axis, which is also the diamond [111] crystallographic axis) and $\{X,Y,Z\}$ is the microscope coordinate system (with $Z$ out of plane for the microscope image).  A $5 \times 5 \times 0.5$ mm$^3$ diamond sample with a 10 $\upmu$m NV layer is illuminated with 1 W of 532 nm laser light, which is passed through an acousto-optic modulator (AOM) switch and flat-top beam shaper. We place the test structures on top of the NV layer, which is coated with 10 nm of Ti metal and 50 nm of Au to prevent light leakage. The diamond is attached to a printed circuit board (PCB) mount that applies a uniform MW magnetic field along the $x$ direction (which is also the lab-frame $X$ direction), and the NVs detect DC magnetic fields (with Ramsey magnetometry \cite{snailPaper}) and AC magnetic fields (with QFM) along the $z$ axis. The mount also acts as a heat sink to dissipate heat deposited by the laser. We applied a $\sim$1 mT bias magnetic field $B_0$ along the $+z$ direction using permanent magnets, which modifies the NV ground-state transition frequencies  through the Zeeman effect. This bias field sets the transition frequencies between NV ground-state sublevels to be $D \pm \gamma B_0$, where $D \approx 2 \pi \times 2870$ MHz is the zero-field splitting and $\gamma \approx 2 \pi \times 28$ GHz/T is the NV gyromagnetic ratio, and we address the $\ket{0} \leftrightarrow \ket{+1}$ transition at $\omega_0 \approx 2 \pi \times2890$ MHz. NV fluorescence light is imaged with an optical microscope onto a digital focal-plane array (DFPA) camera, which has high sensitivity, large dynamic range, 256$\times$256 pixel resolution, on-chip processing, and fast data rates required for this work \cite{dfpaReview}. Laser, MW, test structure current, and camera trigger pulses are programmed with a 1.2 GS/s arbitrary waveform generator (AWG). The camera is triggered to acquire an NV fluorescence readout image across a range of MW pulse durations, and we fit the resulting time traces in each pixel to extract the Rabi frequency for QFM magnetic imaging (Fig.~\ref{fig1drawing}c).

\subsection{Test Structures}
 
\begin{figure*}[t]
\begin{overpic}[width=0.95\textwidth]{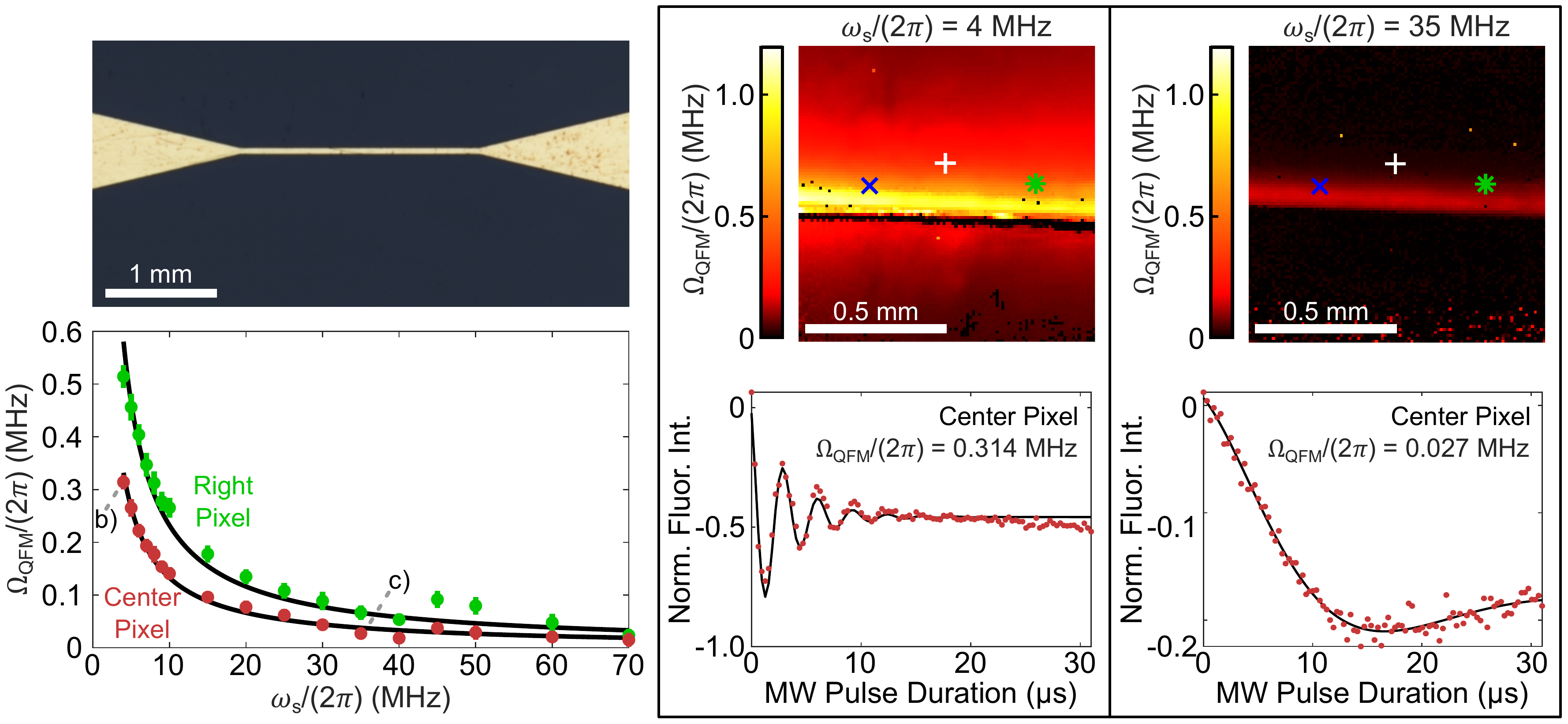}
\centering
\put(0,43){\textsf{\Large a)}}
\put(43,43){\textsf{\Large b)}}
\put(72,43){\textsf{\Large c)}}
\put(0,25){\textsf{\Large d)}}
\end{overpic}
\caption{ (a) Photograph of the straight-wire test structure.  
(b-c) Two QFM imaging examples for an AC current at $\omega_s=2\pi \times 4$ MHz and $2\pi \times 35$ MHz (top). The Rabi oscillations and the extracted Rabi frequency for the center pixel (white cross) are also included (bottom). As expected, $\Omega_{\mathrm{QFM}}$ at a specified pixel decreases as $\omega_{\mathrm{s}}$ increases. 
(d) $\Omega_{\mathrm{QFM}}$ for the center and right (green star) pixels as a function of $\omega_{\mathrm{s}}$ and fit to an $a/\omega_{\mathrm{s}}$ curve, demonstrating the inverse relationship described by Eq.~\ref{eq:qfmConversion}. The pixel marked with a blue $\times$ is discussed in Sec.~\ref{sec:Broken_Assumptions}.
}
\label{fig:StraightWire}
\end{figure*}
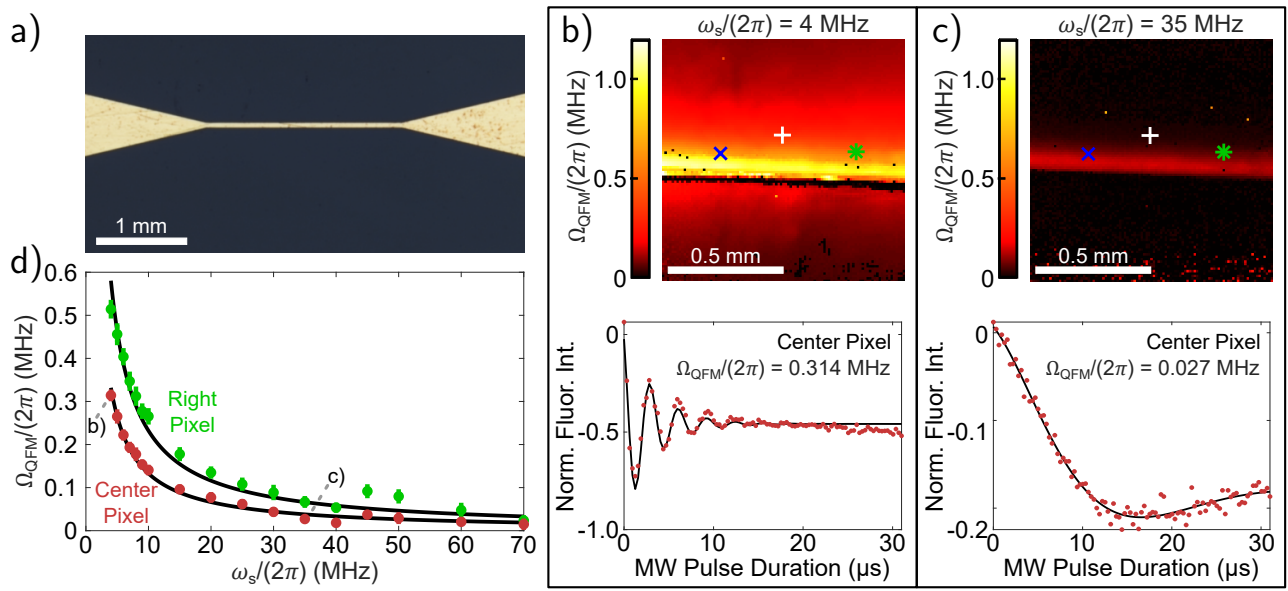

We fabricated two conducting test structures on a glass microscope slide using photolithography and electron-beam evaporation (10 nm Ti, 2 $\upmu$m Al, 10 nm Ti, 150 nm Pt, 200 nm Au), a straight wire and an Archimedean spiral (both with 50 $\upmu$m trace width), as shown in Fig.~\ref{fig:StraightWire}a and Fig.~\ref{fig:Spirals}a. We attached wire leads to the test structures using silver paint (yielding few-$\ohm$ total resistances), after which they were placed on the diamond sample. When applying DC and AC currents using a function generator, we included a 50 $\ohm$ load resistor in series and measured the voltage across it to validate the DC and AC current amplitudes through the test structures. To avoid the possibility of shorting the test structures through the conductive film on the NV layer, we placed a 7.5 $\upmu$m Kapton film between the slide and the diamond.

\subsection{Experimental Procedure}
\label{sec:ExperimentalProc}
Figure \ref{fig:PulseSequences} shows the pulse sequences used for magnetic imaging experiments. First, we perform a conventional Rabi oscillation measurement \cite{maletinskyMWimger} with no current in the test structure (Fig.~\ref{fig:PulseSequences}a).  To do this, we initialize the NVs to the $\ket{0}$ state with a laser pulse, apply an on-resonance MW field with a variable duration $\tau$, and read out the NV final-states with a second laser pulse and the camera. Fitting the NV fluorescence intensity as a function of $\tau$ for each pixel, this yields a map of the on-resonance Rabi frequencies $\Omega_0$ (and implicitly the $\Omega_{\mathrm{b}}$ bias amplitudes), and also the $\pi/2$ pulse duration used for Ramsey magnetometry. 

Next, we apply a DC current to the test structure and perform Ramsey magnetometry (Fig.~\ref{fig:PulseSequences}b). This ensures that the test structure is centered in the camera field of view and allows us to confirm that the test structure is flat on the diamond surface.  We use double-quantum Ramsey spectroscopy for improved magnetic sensitivity and to avoid frequency shifts due to temperature and strain inhomogeneity \cite{snailPaper, turtlePaper}. After laser initialization, the first $\pi/2$-pulse places the NVs in a superposition on the equator of the Bloch sphere, and they accumulate a phase $\phi = 2 \gamma \Delta B \tau$, where $\Delta B$ is the field from the test structure and $\tau$ is the phase accumulation time, then apply a second $\pi/2$-pulse and a readout laser pulse.  After optimizing the Ramsey imaging measurement parameters, we image the DC magnetic field map from the test structure. This provides a DC magnetic image to compare with subsequent AC QFM magnetic images.

Finally, we apply an AC current at frequency $\omega_{\mathrm{s}}$ to the test structure, a MW field at $\omega_{\mathrm{b}} = \omega_0 + \omega_{\mathrm{s}}$, and measure QFM Rabi oscillations (Fig.~\ref{fig:PulseSequences}c). The QFM experiment protocol is nearly identical to the Rabi oscillation experiment protocol, with the main difference being that we address the NVs with frequency $\omega_{\mathrm{b}}$ instead of $\omega_0$. Fitting the fluorescence time-trace data to extract $\Omega_{\mathrm{QFM}}$, we can then use Eq.~\ref{eq:qfmConversion} to get the AC magnetic field amplitude for each pixel ($B_{\mathrm{applied}} = \Omega_{\mathrm{sz}} / \gamma$).

To improve the long-term stability and pixel-to-pixel noise floor of our apparatus, each of the above experiments modulates the signal amplitude between alternate camera exposures (either on/off for Rabi and QFM or positive/negative DC current for Ramsey) \cite{QDM1ggg, maletinskyMWimger, turtlePaper}.

\section{Results}

\subsection{Validation with a Straight Wire Test Structure}

Figure \ref{fig:StraightWire}a shows a photograph of the straight-wire test structure.  Following the measurement procedure described above, we first imaged Rabi oscillations in the field of view with no current in the wire, then measured a 1 mA DC current through the wire to check its position in the field of view. The DC magnetic image yields a conservative estimate that the standoff distance between the NV layer and the wire is $\sim$50 $\upmu$m.

For the QFM magnetic imaging measurement, we applied an 80 mA peak-amplitude AC current through the test structure and measured the projection of the created magnetic field along the $z$-axis. The acquisition time for each $\omega_{\mathrm{s}}$ image was $\sim$5 minutes, roughly half of which was dead time due to camera transfer lag \cite{suppl}.  The Rabi frequency $\Omega_{\mathrm{QFM}}$ in each pixel was extracted using the fit function $C(\tau) = A \cos \left(\Omega_{\mathrm{QFM}}\tau+\phi\right) e^{-{\tau}/{T_\mathrm{Rabi}}} + C_0    $ where $C$ is the normalized fluorescence intensity \cite{suppl}, $A$ is the Rabi contrast amplitude, $\phi$ is a phase offset, $T_\mathrm{Rabi}$ is the Rabi coherence time, and $C_0$ is an offset. Figure \ref{fig:StraightWire}b-c maps the $\Omega_{\mathrm{QFM}}$ values for $\omega_{\mathrm{s}} = 2\pi \times 4$ MHz and 35 MHz, including the QFM Rabi oscillations for the center (white cross) pixel. The black stripe corresponding to $\Omega_{\mathrm{QFM}}$ near 0 is where the wire magnetic field is perpendicular to the $z$ direction.  

Figure \ref{fig:StraightWire}d plots $\Omega_{\mathrm{QFM}}$ as a function of $\omega_{\mathrm{s}}$ for two pixels. As expected, $\Omega_{\mathrm{QFM}}$ decreases as $\omega_{\mathrm{s}}$ increases.  These data are fitted with the fit function $\Omega_{\mathrm{QFM}} = a / \omega_{\mathrm{s}}$ (where $a$ is a free parameter), and are consistent with Eq.~\ref{eq:qfmConversion}. For both pixels, $\Omega_{\mathrm{b}}$ was experimentally determined to be $\sim$2$\pi \times 1.0$ MHz by measuring the on-resonance Rabi oscillation frequencies (Fig.~\ref{fig:PulseSequences}a) and assuming the MW amplitude is the same at $\omega_{\mathrm{b}}$. $\Omega_{\mathrm{sz}}$ is estimated to be $\sim$2$\pi \times 1.8$ MHz and $\sim$3.8 MHz for the center and right (green star) pixels, respectively, based on the simulated magnetic field of a 50 $\upmu$m ribbon at $\sim$50 $\upmu$m standoff distance. The fitted $a$ coefficients are $(2\pi)^2 \times \{1.3,2.3\}$ MHz$^2$ for the center and right pixels, and given $\Omega_{\mathrm{b}} \approx 2\pi \times 1.0$ MHz for both pixels, this yields $\Omega_{\mathrm{sz}} \approx 2\pi \times \{1.3,2.3\}$ MHz. These extracted values are reasonably consistent with the estimated $\Omega_{\mathrm{sz}} \approx 2\pi \times \{1.8,3.8 \} $ MHz.  

Previously, QFM Rabi magnetometry was demonstrated for a bulk (single-pixel) NV ensemble \cite{guoqingQFM}. The above straight-wire test structure demonstration shows that this technique can also be applied to an NV magnetic imaging apparatus, and validates the expected frequency-scaling law predicted in Eq~\ref{eq:qfmConversion}.

\begin{figure*}[hbt!]
\begin{overpic}[width=0.95\textwidth]{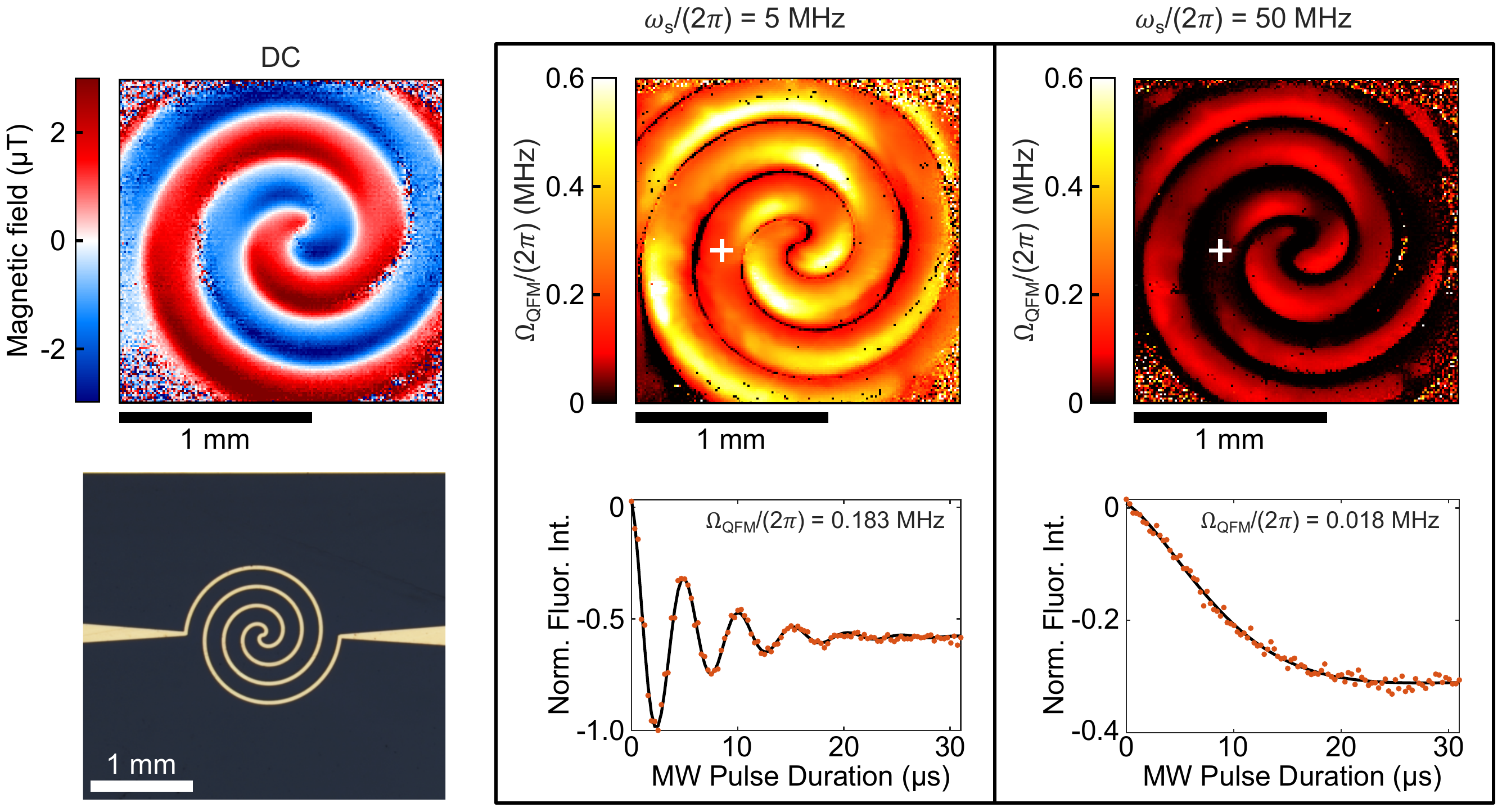}
\centering
\put(1,48){\textsf{\Large b)}}
\put(1,21){\textsf{\Large a)}}
\put(33,52.5){\textsf{\Large c)}}
\put(66.5,52.5){\textsf{\Large d)}}
\end{overpic}
\caption{ (a)  Photograph of the Archimedean spiral test structure. 
(b) DC magnetic image of the spiral with 1 mA signal current measured using a Ramsey sequence. 
(c-d) Two examples of QFM imaging (top), with the extracted $\Omega_{\mathrm{QFM}}$ for an example pixel (bottom). $\Omega_{\mathrm{QFM}}$ for this pixel decreases by a factor $1/\omega_{\mathrm{s}}$ as the input signal frequency increases from $\omega_s=2\pi \times 5$ MHz to $2\pi \times 50$ MHz.
}
\label{fig:Spirals}
\end{figure*}
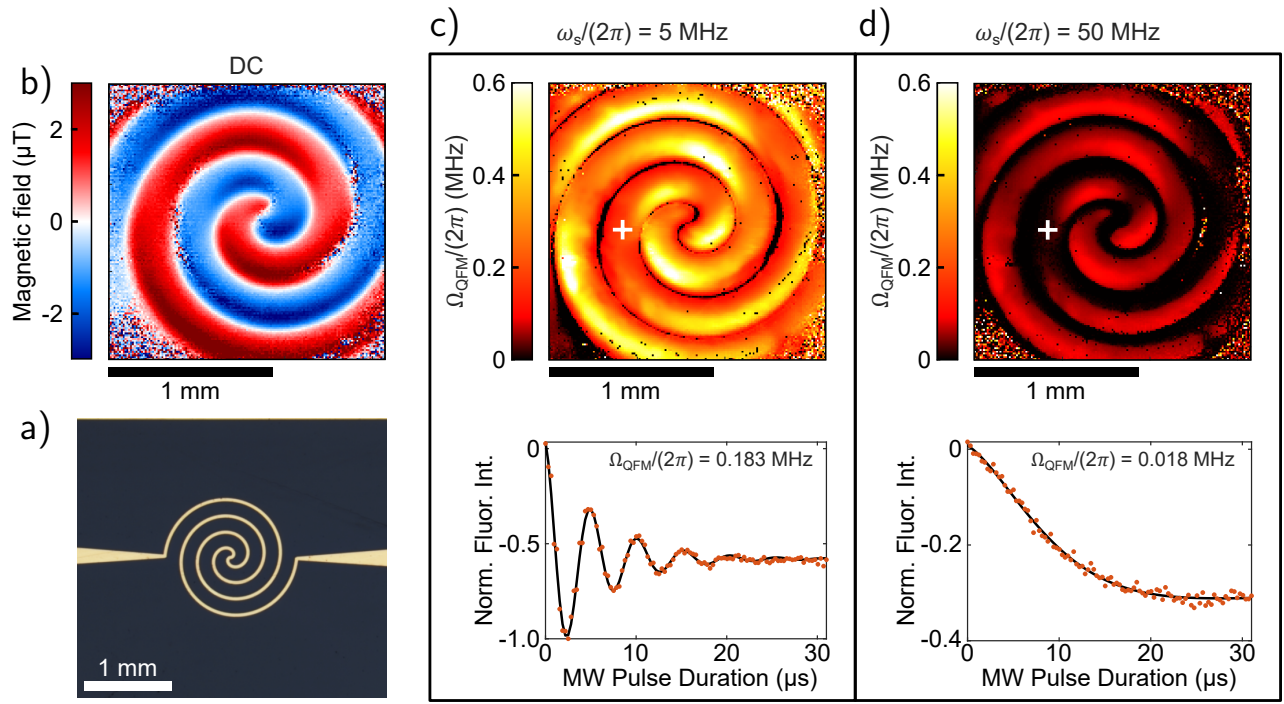

\subsection{Archimedean Spiral Test Structure}
Having demonstrated that the QFM imaging method works as expected for a simple test case, we now apply it to a more complicated test structure, an Archimedean spiral (Fig.~\ref{fig:Spirals}a). Unlike the straight-wire, the spiral test structure has currents (magnetic fields) pointing in a wide range of directions, leading to a nontrivial magnetic field map. The test structure layout is similar to a planar spiral inductor, and generalizing the measurement technique to spiral inductors and other similar passive RF circuit elements is straightforward. Furthermore, since the spiral is a passive circuit element, we control the frequency and amplitude of current passing through it. By contrast, when measuring the magnetic fields from a more sophisticated IC, these parameters are likely uncontrollable (and are often unknown).

Figures \ref{fig:Spirals}b-c show QFM magnetic images for $\omega_{\mathrm{s}} = 2\pi \times 5$ MHz and 50 MHz, with 80 mA peak-amplitude current.  Rabi oscillation data and fits for an example pixel are included, which also show the $\Omega_{\mathrm{QFM}} \propto 1/\omega_{\mathrm{s}}$ dependence expected from Eq.~\ref{eq:qfmConversion}. Figure \ref{fig:Spirals}d includes a DC magnetic image for comparison, acquired using Ramsey magnetometry and showing a similar magnetic field amplitude structure.  Note that the DC experiment measures both positive and negative magnetic field values while the AC experiments measure amplitudes, though the absolute value of the DC image is consistent with the AC images.

\subsection{Large-$\Omega_{\mathrm{sz}}$ Small-$\omega_{\mathrm{s}}$ Regimes}
In Fig.~\ref{fig:StraightWire} and Fig.~\ref{fig:Spirals}, we used Eq.~\ref{eq:qfmConversion} to confirm that the measured QFM Rabi frequencies $\Omega_{\mathrm{QFM}}$ were behaving as expected. However, this expression is valid when
\begin{equation}
    \Omega_{\mathrm{b}},\Omega_{\mathrm{sz}} \ll \omega_{\mathrm{s}},
    \label{eqn:assumptions}
\end{equation}
which is not necessarily true for all experimental conditions. For large $\Omega_{\mathrm{sz}}$ and small $\omega_\mathrm{s}$, $\Omega_{\mathrm{QFM}}$ is more accurately described by
\begin{equation}
    \Omega_{\mathrm{QFM}}=\Omega_\mathrm{b} J_1\left(\frac{2\Omega_{\mathrm{sz}}}{\omega_\mathrm{s}}\right),
\label{eq:Bessel}
\end{equation}
where $J_1$ is the Bessel function of the first kind \cite{suppl}.

\begin{figure}[hbt!]
\begin{overpic}[width=0.4\textwidth]{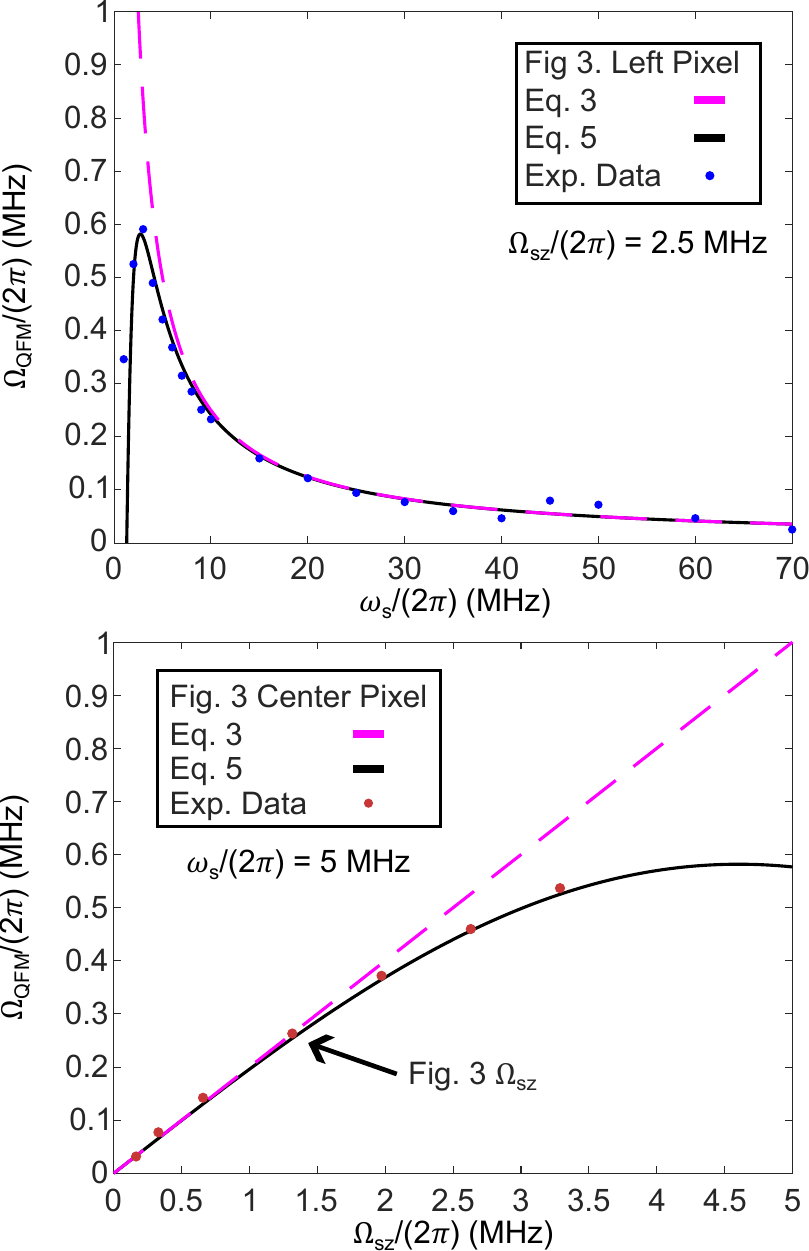}
\centering
\end{overpic}
\caption{(a) $\Omega_{\mathrm{QFM}}$ as a function of $\omega_{\mathrm{s}}$ with a fixed $\Omega_{\mathrm{sz}}$ for the left pixel in Fig.~\ref{fig:StraightWire}. The dashed pink line shows the expected $\Omega_{\mathrm{QFM}}$ predicted by Eq.~\ref{eq:qfmConversion}. The black line shows the expected $\Omega_{\mathrm{QFM}}$ from Eq.~\ref{eq:Bessel}, and the blue dots are the experimental data. 
(b) $\Omega_{\mathrm{QFM}}$ as a function of $\Omega_{\mathrm{sz}}$ at a fixed $\omega_{\mathrm{s}}$ for the center pixel in Fig.~\ref{fig:StraightWire}.
}
\label{fig:Broken_Assumptions}
\end{figure}

Figure \ref{fig:Broken_Assumptions}a plots $\Omega_{\mathrm{QFM}}$ as a function of $\omega_{\mathrm{\mathrm{s}}}$ for the left (blue $\times$) pixel in Fig.~\ref{fig:StraightWire}, showing the necessity of Eq.~\ref{eq:Bessel}. This pixel has a larger $\Omega_{\mathrm{sz}} = 2\pi \times 2.5$ MHz amplitude than the right pixel because of its proximity to the wire, which we estimated using the $\omega_{\mathrm{s}} \ge 2\pi \times 30$ MHz tail and Eq.~\ref{eq:qfmConversion}. We also measured its $\Omega_{\mathrm{QFM}}$ frequencies down to a lower-bound $\omega_{s} = 2\pi \times 2$ MHz. These $\Omega_{\mathrm{QFM}}$ frequencies were measured with the same 80 mA peak-amplitude AC current as in Fig.~\ref{fig:StraightWire}.  Figure \ref{fig:Broken_Assumptions}a  verifies that Eq.~\ref{eq:Bessel} (solid black line) is a more appropriate description of the $\Omega_{\mathrm{QFM}}$ behavior for small $\omega_{\mathrm{s}}$ than Eq.~\ref{eq:qfmConversion} (blue dashed line).

Figure \ref{fig:Broken_Assumptions}b shows $\Omega_{\mathrm{QFM}}$ as a function of $\Omega_{\mathrm{sz}}$ (i.e.~applied current) for the center pixel previously examined in Fig.~\ref{fig:StraightWire}, using $\omega_{\mathrm{s}} = 2\pi \times 5$ MHz. For sufficiently large $\Omega_{\mathrm{sz}}$, the measured $\Omega_{\mathrm{QFM}}$ deviates from the linear dependence predicted by Eq.~\ref{eq:qfmConversion} (blue dashed line), and is more accurately described by Eq.~\ref{eq:Bessel} (solid black line).

\label{sec:Broken_Assumptions}

\section{Discussion}

Our analysis shows the dynamic range of the QFM AC magnetometry and imaging method. In the large-$\Omega_{\mathrm{sz}}$ and small-$\omega_\mathrm{s}$ regime, conversion from $\Omega_{\mathrm{QFM}}$ to $\Omega_{\mathrm{sz}}$ (which is straightforward using Eq.~\ref{eq:qfmConversion}) requires inverting Eq.~\ref{eq:Bessel}. $\Omega_{\mathrm{QFM}}$ does not scale monotonically with $\omega_{\mathrm{s}}$ and $\Omega_{\mathrm{sz}}$, and near a maximum in $\Omega_{\mathrm{QFM}}$, $\Omega_{\mathrm{QFM}}$ is insensitive to $\Omega_{\mathrm{sz}}$. These details may complicate the QFM AC magnetometry method when applying it to more general experimental situations.

Equation \ref{eq:qfmConversion} implies that one can increase $\Omega_{\mathrm{b}}$ to compensate for a small $\Omega_{\mathrm{sz}}$, since a slow $\Omega_{\mathrm{QFM}}$ is difficult to measure if its period is slow compared to the coherence time $T_\mathrm{Rabi}$. However, this may also require additional analysis to interpret the measured $\Omega_{\mathrm{QFM}}$ values \cite{suppl}. Alternatively, in a regime where $\Omega_{\mathrm{QFM}}$ is slow compared to the $T_{\mathrm{Rabi}}$ coherence time, one can apply an on-resonance $\pi/2$-pulse at frequency $\omega_0$ before the QFM Rabi pulse at frequency $\omega_{\mathrm{b}}$, making the NV fluorescence intensity proportional to $\sin \left(\Omega_{\mathrm{QFM}}\tau+\phi\right) e^{-{\tau}/{T_\mathrm{Rabi} }}$ instead of $\cos \left(\Omega_{\mathrm{QFM}}\tau+\phi\right) e^{-{\tau}/{T_\mathrm{Rabi}}}$ \cite{scott_pT_RabiMag}.  For small $\Omega_{\mathrm{QFM}}$, the fluorescence change will be proportional to $\Omega_{\mathrm{QFM}} \tau$ (sometimes called ``sine magnetometry") instead of $(\Omega_{\mathrm{QFM}} \tau)^2$ (``cosine magnetometry") for improved sensitivity and minimum-detectable $\Omega_{\mathrm{QFM}}$, though implementing this also requires careful calibration \cite{rmpQuantumSensingReview}.

\section{Conclusion and outlook}

In this work we demonstrated an NV wide-field magnetic imaging apparatus capable of measuring AC magnetic fields using QFM up to 70 MHz. We implemented a Rabi-based magnetometry scheme with linearly-polarized signal and bias fields to measure a signal-field component oscillating along NV $z$-axis.  The imager itself overcomes several of the common challenges for NV pulsed magnetic imaging, achieving sufficient homogeneity to realize a $\sim$1.5$\times$1.5 mm$^2$ FOV. After validating that the technique works with a simple straight-wire test structure, we then imaged an Archimedean spiral test structure to show that this technique can be generalized to spiral inductors and other passive RF circuit elements.  Additional extensions could use spin-locking and pulsed dynamical decoupling methods (such as CPMG), or use circularly-polarized signal and bias fields to measure a signal field oscillating transverse to the NV $z$-axis  \cite{guoqingQFM}. 

Building on previous demonstrations using NV magnetic imaging of DC currents for electronics interrogation and failure analysis (FA) \cite{nvTIVA, NVionTrapFA}, our QFM magnetic imaging approach could also be applied to FA, fabrication process monitoring, and non-contact wafer testing. This would generalize the technique to include AC magnetic field sources at arbitrary frequencies.  One could measure the frequencies and amplitudes of currents in high-speed test structures (e.g.~ring oscillators \cite{qdmFPGA}), which are included on dice or wafers for fabrication process troubleshooting. To achieve this, one could use QFM magnetic imaging to implement a spectrum-analyzer imager (a form of hyperspectral imaging), generating image for a range of $\omega_{\mathrm{s}}$ values \cite{aprelScanMS, hyperspectralReview}. The resulting hyperspectral data cube could then be analyzed for PSA applications, such as counterfeit detection, component aging, and fabrication process variation over time or across a wafer \cite{paiPSA_istfa, specAnaNote}.  This could also be useful for validating how internal current paths within a device change as a function of frequency and impedance (for example, high-frequency capacitive shorting). 

As a further extension, our QFM magnetic imaging approach could be adapted for space-domain reflectometry for open-circuit fault localization. This was previously demonstrated with a scanning RF SQUID sensor  \cite{neoceraSpaceDomainRefl}, but could potentially achieve improved spatial resolution and sensitivity with an NV magnetic microscopy apparatus. Finally, this technique could be used to evaluate the RF and MW images of hardware used for quantum computing, such as ion trap chips and superconducting qubits. We note the emerging synergy between quantum technologies, and here one quantum technology (quantum sensing and imaging) would troubleshoot hardware of another technology (quantum computing). 

\section{Acknowledgements}
We thank M.~Ricci for help with test structure fabrication and packaging, C.~Reed for help with preparing Fig.~\ref{fig1drawing}a, and M.~Jordan, C.~Nordquist, and D.~Krawczyk for useful discussions.


\begin{thebibliography}{27}
\bibitem [{spe()}]{specAnaNote}%
  \BibitemOpen
  \href@noop {} {}\bibinfo {note} {We note that previous works have also
  combined NV magnetic imaging with a spectrum analyzer \cite{spatialMWspec,
  spatialMWspec2}, but in these cases the imager itself is also the spectrum
  analyzer.}\BibitemShut {Stop}%
\end{thebibliography}

%

\clearpage


\section*{Supplementary material}

# Supplemental Material for "Quantum Frequency Mixing using an NV Diamond Microscope"

Samuel J. Karlson,[1,2,3] Pauli Kehayias,[1] Jennifer M. Schloss,[1] Andrew C. Maccabe,[1,*]
David F. Phillips,[1] Guoqing Wang,[2,3,4] Paola Cappellaro,[2,3,4] and Danielle A. Braje[1]

[1]*MIT Lincoln Laboratory, Lexington, MA 02421, USA*
[2]*Department of Nuclear Science and Engineering, Massachusetts Institute of Technology, Cambridge, MA 02139, USA*
[3]*Research Laboratory of Electronics, Massachusetts Institute of Technology, Cambridge, MA 02139, USA*
[4]*Department of Physics, Massachusetts Institute of Technology, Cambridge, MA 02139, USA*

## I. EXPERIMENT DEAD-TIME ASSESSMENT

For an experiment measuring QFM Rabi oscillations (Fig. 2c in the main text), each pulse sequence takes 125 µs. The pulse sequence includes a 25 µs laser initialization pulse, a MW pulse ranging from $\tau = 50$ ns to 31 µs, and a 8 µs fluorescence readout laser pulse, with the total pulse sequence lasting 65 µs. For each MW pulse duration, the sequence repeats for 32 camera exposures (125 µs × 32 = 4 ms), and the fluorescence counts are summed and off-loaded from the camera to the PC frame grabber card as one "frame", taking 0.5 ms to off-load. This corresponds to 4.5 ms spent acquiring per MW pulse duration.

This procedure is repeated for each of the 200 MW pulse durations used, where every other time point disables the bias field (sets $\Omega_b = 0$) for background subtraction. This yields two stacks of 100 images, which take 0.9 s to acquire. Additional software lag extends this to ∼1.5 s per "capture" of a full stack of QFM Rabi oscillation images. Of this 1.5 s, roughly 0.1 s (camera transfer) + 0.6 s (software lag) = 0.7 s is dead time, though definitions of "dead time" may also vary.

The total acquisition time for the time-averaged images in Figs. 3 and 4 in the main text is ∼5 minutes (200 captures).

## II. FLUORESCENCE NORMALIZATION

NV fluorescence intensities are often normalized to the $|0\rangle$ state fluorescence intensity. We present our results differently because the DFPA camera uses on-chip subtraction, reporting intensity differences, not the original intensities [S1]. Fortunately, since we extract Rabi frequencies from the time traces of each pixel, the units of the vertical scale (e.g. photoelectrons, volts, or normalized dimensionless units) themselves are not critical. However, to provide a means for comparing the fluorescence intensity contrast between pixels and measurements, the fluorescence intensities in Figs. 3 and 4 in the main text are scaled to the minimum and maximum fluorescence intensities of the selected example pixel in Fig. 4 (white cross). The fluorescence intensities at $\tau = 0$ are not modified, and the minimum fluorescence intensity in the lower sub-figure of Fig. 4b is used as the normalization value.

## III. MAGNETIC NOISE FLOOR ASSESSMENT

Here we estimate the pixel-to-pixel noise floor [S2, S3] for AC magnetic imaging using the QFM method. We can solve Eq. 3 in the main text to convert a measured $\Omega_{\text{QFM}}$ to an AC magnetic field amplitude $\Omega_{\text{sz}}$,

$$\Omega_{\text{sz}} = \frac{\omega_s}{\Omega_b} \Omega_{\text{QFM}}. \tag{S1}$$

If $\Omega_b$ is homogeneous and is known to a small uncertainty before the QFM measurement, and $\omega_s$ is fixed, then the pixel-to-pixel noise floor in $\Omega_{\text{sz}}$ is

$$\delta\Omega_{\text{sz}} = \frac{\omega_s}{\Omega_b} \delta\Omega_{\text{QFM}}. \tag{S2}$$

---

[*] Current Affiliation: Quantum Science and Engineering Program, Harvard University, Cambridge, MA 02138, USA

Ideally $\delta\Omega_{\text{QFM}}$ is determined by the photon shot noise of the NV fluorescence light imaged onto each camera pixel, which can be evaluated experimentally (keeping track of measurement repetition rate and dead time) and compared to the expected theoretical value.

### IV. QFM RABI FREQUENCY IN THE LARGE-$\Omega_{\text{sz}}$ SMALL-$\omega_{\text{s}}$ REGIMES

The system Hamiltonian is

$$H = \frac{\omega_0}{2}\sigma_z + \Omega_{\text{sz}}\sigma_z \cos(\omega_{\text{s}}t) + \Omega_{\text{b}}\sigma_x \cos(\omega_{\text{b}}t). \tag{S3}$$

We can go into a first rotating frame at the $\omega_{\text{b}}$ frequency:

$$H_2 = \frac{\sigma_z}{2}[2\Omega_{\text{sz}}\cos(\omega_{\text{s}}t) - (\omega_{\text{b}} - \omega_0)] + \frac{\Omega_{\text{b}}}{2}\sigma_x. \tag{S4}$$

We neglect the counter-rotating term at $2\omega_{\text{b}}$. This is valid if $\omega_{\text{b}} \gg |\omega_0 - \omega_{\text{b}}|$, which is the case here (but the approximation still fails for large $\Omega_{\text{b}}$.) Note that $(\omega_{\text{b}} - \omega_0) = \omega_{\text{s}} + \delta_z$, with $\delta_z \approx 0$.

We can define yet another rotating frame set by the $\sigma_z$ term in $H_2$, $H_2^{(z)}$. The rotating frame transformation is given by the unitary

$$U_3 = \exp\left(-i\int_0^t H_2^{(z)}(t')dt'\right) \equiv \exp(-i\phi\sigma_z/2)$$
$$= \exp\left(-i\frac{\sigma_z}{2}[\frac{2\Omega_{\text{sz}}}{\omega_{\text{s}}}\sin(\omega_{\text{s}}t) - (\omega_{\text{s}} + \delta_z)t]\right). \tag{S5}$$

Since $e^{i\phi\sigma_z/2}\sigma^+ e^{-i\phi\sigma_z/2} = e^{-i\phi}\sigma^+$, we want to calculate

$$\exp\left(-i[\frac{\Omega_{\text{sz}}}{\omega_{\text{s}}}\sin(\omega_{\text{s}}t) - (\omega_{\text{s}} + \delta_z)t]\right) \tag{S6}$$

$$= \sum_{n=0}^{\infty} J_n(2\Omega_{\text{sz}}/\omega_{\text{s}})e^{-i[n\omega_{\text{s}} - (\omega_{\text{s}} + \delta_z)]}, \tag{S7}$$

where $J_n$ are Bessel functions of the first kind. The Hamiltonian in the rotating frame, $H_3 = U_3^\dagger H_2 U_3 - H_2^{(z)}$, is then

$$H_3 = \frac{\Omega_{\text{b}}}{2}\sum_n J_n(2\Omega_{\text{sz}}/\omega_{\text{s}})(\sigma_x \cos[n\omega_{\text{s}} - (\omega_{\text{s}} + \delta_z)]t \tag{S8}$$
$$+ \sigma_y \sin[n\omega_{\text{s}} - (\omega_{\text{s}} + \delta_z)]t).$$

The dominant time-independent term for $\delta_z \approx 0$ is $n = 1$. Note that instead for $\omega_{\text{s}}$ relatively large we do not have just a single dominating term, and thus this approach does not give a good solution. Here instead, this approximation gives an amplitude

$$\Omega_{\text{QFM}} = \Omega_{\text{b}} J_1\left(\frac{2\Omega_{\text{sz}}}{\omega_{\text{s}}}\right), \tag{S9}$$

which reproduces the data shown in Fig. 5 of the main text.

#### A. Corrections

We can consider the lowest-order correction for $H_3$ at $\delta_z = 0$:

$$H_3 \approx \frac{\Omega_{\text{b}}}{2}\left[J_1(2\Omega_{\text{sz}}/\omega_{\text{s}})\sigma_x\right.$$
$$\left.+ J_0(2\Omega_{\text{sz}}/\omega_{\text{s}})(\sigma_x \cos(\omega_{\text{s}}t) - \sigma_y \sin(\omega_{\text{s}}t)\right]. \tag{S10}$$

Note that for $2\Omega_{sz} < \omega_s$ we have $J_1 < J_0$, thus the oscillating term might dominate. While this looks similar to Eq. S4, here $\omega_s < \Omega_b J_1(2\Omega_{sz}/\omega_s)$, so we cannot take the same route. We can instead use the Floquet expansion used in [S4]. We then obtain the following detuning as the leading correction,

$$H_3 \approx \frac{\Omega_b}{2} J_1(2\Omega_{sz}/\omega_s)\sigma_x + \frac{\Omega_b^2 J_0^2(2\Omega_{sz}/\omega_s)}{4\omega_s}\sigma_z. \tag{S11}$$

---



\end{document}